\documentclass[aps,prd,onecolumn,twoside,notitlepage,preprintnumbers,showpacs,
showkeys,superscriptaddress,byrevtex,floatfix,nofootinbib,amsmath,amssymb,longbibliography]{revtex4-1}

\usepackage{graphicx}% Include figure files
\usepackage{dcolumn}% Align table columns on decimal point
\usepackage{bm}% bold math
\usepackage{tensor}
\usepackage{cancel}
\usepackage{array,amscd,amsmath,amssymb,amsthm,graphicx,mathtools,placeins,slashed,verbatim,tensor,dsfont}
\usepackage[bookmarksnumbered,bookmarksopen=true,bookmarksopenlevel=1,pdfstartview=FitH,hidelinks]{hyperref}
\usepackage[all]{hypcap}
\usepackage{rotating}
\usepackage[caption=false]{subfig}
\usepackage{eepic}

\newcommand{\R}{\mathds{R}}
\newcommand{\C}{\mathds{C}}
\newcommand{\Q}{\mathds{H}}
\newcommand{\Oct}{\mathds{O}}

\newcommand{\g}{\gamma}
\newcommand{\ghat}{\hat\gamma}
\newcommand{\N}{\mathcal{N}}
\newcommand{\F}{\mathds{F}}
\newcommand{\Ftilde}{\widetilde{\mathds{F}}}

\newcolumntype{L}[1]{>{\raggedright\let\newline\\\arraybackslash\hspace{0pt}}m{#1}}
\newcolumntype{C}[1]{>{\centering\let\newline\\\arraybackslash\hspace{0pt}}m{#1}}
\newcolumntype{R}[1]{>{\raggedleft\let\newline\\\arraybackslash\hspace{0pt}}m{#1}}

\DeclareMathOperator{\SO}{SO}

\DeclareMathOperator{\SL}{SL}
\DeclareMathOperator{\SU}{SU}

\newcommand{\be}{\begin{equation}}
\newcommand{\ee}{\end{equation}}

\begin{document}

\preprint{Imperial/TP/2015}

\title{Octonionic $D=11$ Supergravity \\
and `Octavian Integers' as Dilaton Vectors} 

\author{A. Anastasiou}
\email[]{alexandros.anastasiou07@imperial.ac.uk}
\affiliation{Theoretical Physics, Blackett Laboratory, Imperial College London,
London SW7 2AZ, United Kingdom}
\author{M. J. Hughes}
\email[]{mia.hughes07@imperial.ac.uk}
\affiliation{Theoretical Physics, Blackett Laboratory, Imperial College London,
London SW7 2AZ, United Kingdom}

\date{\today}

\begin{abstract}
We formulate $D=11$ supergravity over the octonions by rewriting 32-component Majorana spinors as 4-component octonionic spinors. Dimensional reduction to $D=4$ and $D=3$ suggests an interpretation of the so-called `dilaton vectors', which parameterise the couplings of the dilatons to other fields in the theory, as unit `octavian integers' -- the octonionic analogues of integers. The parameterisation involves a novel use of the duality between points and lines on the Fano plane, and suggests a series of consistent truncations with $\N=8\rightarrow4\rightarrow2\rightarrow1$, giving the `four curious supergravities' studied by Duff and Ferrara.
\end{abstract}

\pacs{11.25.Yb, 11.30.Pb, 11.25.Mj}

\keywords{M-Theory, Supergravity, Octonions}

\maketitle
%\tableofcontents

\section{Introduction}

Of the four normed division algebras -- the real numbers $\R$, the complex numbers $\C$, the quaternions $\Q$ and the octonions $\Oct$ -- the fourth is perhaps the most intriguing and the most mysterious. The octonions themselves hold an exceptional status, as well as providing an intuitive language for describing the exceptional groups, which appear as various symmetries in string and M-theory. The connection between supersymmetry, string theory and the division algebras has been studied in a few different contexts over the years. See, for example, \cite{Kugo:1982bn,Duff:1987qa,Manogue:1993ja,
Baez:2009xt,Baez:2010ye,Evans:1994,Borsten:2008wd,Borsten:2013bp,Anastasiou:2013cya,Rooman:1983es,Anastasiou:2013hba}. However, it is fair to say that the full significance of the octonions in string theory remains puzzling.

In the present paper, in an attempt to shed light on this problem, we present the Lagrangian and transformation rules of $D=11$ supergravity written over the octonions. Following \cite{Anastasiou:2014zfa}, the method relies on the fact that a $D=11$ spinor with 32 components may be packaged as a 4-component octonionic column vector \cite{Baez:2010ye,Toppan:2003ry,Anastasiou:2014zfa}. We consider dimensional reduction to $D=4$ and $D=3$, where the U-duality groups are $E_{7(7)}$ and $E_{8(8)}$, respectively. The coupling of the 7 or 8 dilatons to the other scalar fields in the theory can be parameterised by the sets of $E_{7(7)}$ or $E_{8(8)}$ root vectors \cite{Lu:1995yn,PhysRevD.83.046007}. The octonionic nature of the fields in the Lagrangian suggests a new perspective in which these root vectors, or so-called `dilaton vectors', are unit-norm `octavian integers' \cite{coxeter1946} -- the octonionic analogues of the integers. This  involves a novel use of the dual Fano plane, which is obtained by interchanging points and lines on the Fano plane. We demonstrate how in $D=4$ our parameterisation suggests a simple series of truncations with $\N=8\rightarrow4\rightarrow2\rightarrow1$, giving the so-called `four curious supergravities' studied by Duff and Ferrara in \cite{PhysRevD.83.046007}.

\section{The Octonions}

\begin{figure}[h!]
  \centering
    \includegraphics[width=0.3\textwidth]{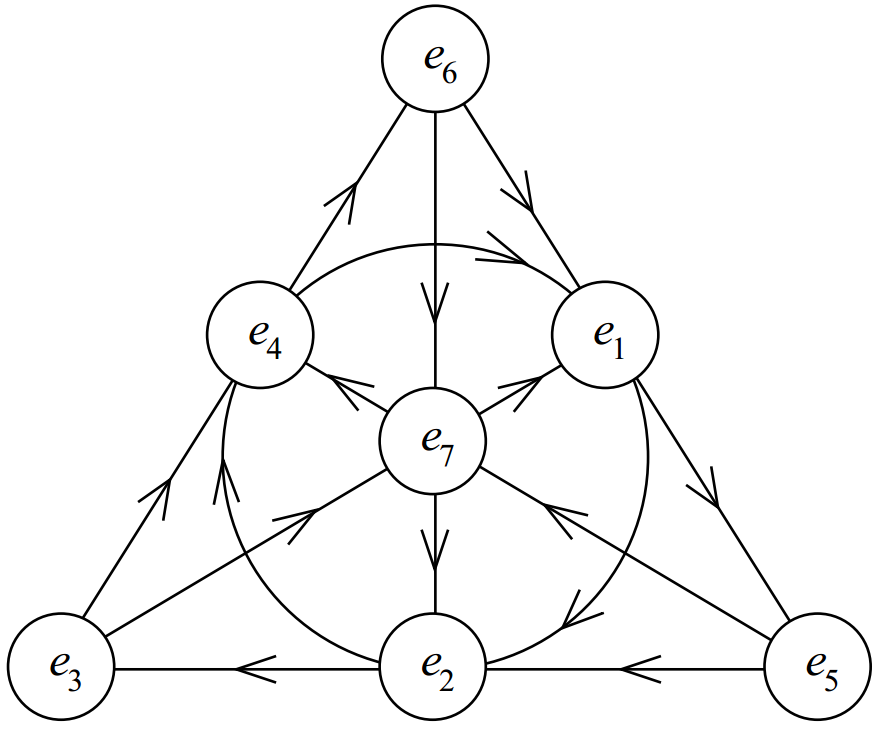}
  \caption{\footnotesize{The Fano plane $\F$ (image from \cite{Baez:2001dm}). Each oriented
line corresponds to a quaternionic subalgebra. For example, $e_2e_3=e_5$
and cyclic permutations; odd permutations go against the direction of the
arrows on the Fano plane and we pick up a minus sign, e.g. $e_3e_2=-e_5$.}}\label{FANO}
\end{figure}

In this section, we briefly introduce the basic properties of the octonions. The octonions $\Oct$ are an 8-dimensional, non-commutative, \emph{non-associative} normed division algebra with basis $e_a$, $a=0,\cdots,7$. A general octonion $x\in\Oct$ is then written as the linear combination $x=x_{a}e_{a}$, with $x_a\in\R$. The first basis element $e_0=1$ corresponds to the `real part', while the
other $7$ basis elements $e_i$, where $i=1,\cdots,7$, are `imaginary':
\be
e_0^2=1,~~~~e_i^2=-1.
\ee
We define a linear involution denoted by *, which changes the sign of the imaginary basis elements:
\be
{e_0}^*=e_0,~~~~{e_i}^*=-e_i.
\ee
Using this we can extract the real and imaginary parts of $x\in\Oct$ by
\be
\text{Re}(x)\equiv\frac{1}{2}(x+x^*)=x_0,~~~~~~\text{Im}(x)\equiv\frac{1}{2}(x-x^*)=x_ie_i.
\ee
The multiplication rule for the imaginary octonionic basis elements  is given by
\be\label{IMMULT}
e_ie_j=-\delta_{ij}+C_{ijk}e_k,
\ee
where the totally antisymmetric tensor $C_{ijk}$ is zero unless $ijk$ lie in a line of the Fano plane $\F$ -- see Fig. \ref{FANO}:
\be
\begin{split}
C_{ijk}= 1 \hspace{0.1cm}\text{ if }\hspace{0.1cm}ijk\in\bold{L}=\{124,235,346,457,561,672,713\}.
\end{split}
\ee
Note that any subalgebra of $\Oct$ spanned by $\{e_0,e_i,e_j,e_k\}$ with $ijk\in\bold{L}$
is isomorphic to the quaternions. 

%We can generalise equation \eqref{IMMULT} by including the identity component $e_0=1$:
%\be\label{OCTMULT}
%\begin{split}
%&{e_a}{e_b}=\left(+\delta_{a0}\delta_{bc}+\delta_{0b}\delta_{ac}-\delta_{ab}\delta_{0c}+C_{abc}\right)e_c,\\
%&{e_a^*}{e_b}=\left(+\delta_{a0}\delta_{bc}-\delta_{0b}\delta_{ac}+\delta_{ab}\delta_{0c}-C_{abc}\right)e_c,\\
%&{e_a}{e_b^*}=\left(-\delta_{a0}\delta_{bc}+\delta_{0b}\delta_{ac}+\delta_{ab}\delta_{0c}-C_{abc}\right)e_c,
%\end{split}
%\ee
%where the tensor $C_{abc}$ is also totally antisymmetric with $C_{0ab}=0$. 

%Although the octonions are non-associative, they do exhibit a similar but weaker property
%called \emph{alternativity}. An algebra $\Al$ is alternative if and only if for
%all $x,y\in\Al$ we have:
%\be\label{altern}
%(xx)y=x(xy),~~~~~~(xy)x=x(yx),~~~~~~(yx)x=y(xx)
%\ee
%(note that any one of these three conditions follows from the other two).
%The alternativity condition may also be expressed in terms of a trilinear map
%called the associator given by:
%\be
%[x,y,z]=(xy)z-x(yz),~~~~x,y,z\in\Al,
%\ee
%which measures the failure of associativity. An algebra $\Al$ is alternative
%if and only if the associator is an antisymmetric function of its three arguments, since setting any two of the arguments to be %equal then yields one of the three conditions in (\ref{altern}).

We can define the norm $|\hspace{-0.2mm}|x|\hspace{-0.2mm}|$ of an octonion
$x$ by
\be\label{NORM}
|\hspace{-0.2mm}|x|\hspace{-0.2mm}|^2=xx^*=x^*x=x_a x_a,
\ee
which satisfies the `normed division algebra' property
\be
|\hspace{-0.2mm}|xy|\hspace{-0.2mm}|=|\hspace{-0.2mm}|x|\hspace{-0.2mm}|\,|\hspace{-0.2mm}|y|\hspace{-0.2mm}|.
\ee
Polarising this gives a natural inner product \cite{Baez:2001dm}:
\be\label{INNER}
\langle{x}|{y}\rangle=\frac{1}{2}\left(xy^*+yx^*\right)=\frac{1}{2}\left(x^*y+y^*x\right)=x_ay_a\hspace{0.4cm}\text{i.e.}\hspace{0.4cm}\langle{e_a}|{e_b}\rangle=\delta_{ab}.
\ee

The complement of a line in the Fano plane is called a quadrangle. Thus the Fano plane $\F$ has seven points, seven lines and seven quadrangles. It will be useful to define the set of quadrangles
\be
\bold{Q}=\{3567,4671,5712,6123,7234,1345,2456\}.
\ee
%Just as multiplication of the octonionic basis elements
%is encoded in the lines of the Fano plane, the associator of three octonionic
%basis elements is encoded in its seven quadrangles $\bold{Q}$:
%\be
%[e_a,e_b,e_c]=2Q_{abcd}e_d,
%\ee
%where the tensor $Q_{abcd}$ is totally antisymmetric with $Q_{0abc}=0$, and
%the non-zero $Q_{ijkl}$ are given by:
%\be
%Q_{ijkl}=1\hspace{0.2cm}\text{if}\hspace{0.2cm}ijkl\in\bold{Q}=\{3567,4671,5712,6123,7234,1345,2456\}.
%\ee
%By definition,
%the tensors $Q_{ijkl}$ and $C_{ijk}$ are dual to one another in seven dimensions:
%\be
%Q_{ijkl}=-\frac{1}{3!}\varepsilon_{ijklmnp}C_{mnp}.
%\ee

\section{The Kirmse and Octavian Integers}

In this section we very briefly discuss octonionic number theory. By analogy with the usual set of integers $\mathds{Z}\subset\R$, an octonionic integer system $\mathds{I}$ should be an 8-dimensional lattice embedded in $\Oct$, which is (preferably) closed under multiplication using the rule inherited from $\Oct$. The most obvious example is of course to take octonions whose components are all integers:
\be
\mathds{I}=\{x=x_a e_a\in\Oct~|~x_a\in\mathds{Z}\}.
\ee
However, as shown in \cite{Conway}, for a richer number theory we can require an analogue of the unique prime factorisation theorem to hold in $\mathds{I}$. For the ordinary set of integers $\mathds{Z}$ this theorem says that each integer is a product of positive or negative primes in a way that is unique up to order and sign change. For an analogue of this theorem to hold in $\mathds{I}$, it must be `well-packed' \cite{Conway}; that is, the following two conditions must hold:
\begin{enumerate}
\item{there is no element of $\Oct$ whose distance is $\geq{1}$ from the nearest lattice point of $\mathds{I}$},
\item{the distance between any lattice point and any other lattice point is $\geq{1}$},
\end{enumerate}
where the distances are evaluated using the norm in $\Oct$ -- see equation \eqref{NORM}.

One set of octonions that satisfies these two conditions is the so-called \emph{Kirmse integers} $\mathds{K}$. These can be described as follows. We can always write an octonion $x$ as
\be
x=(x_0+x_ie_i +x_je_j+x_ke_k)+(x_{i'}e_{i'}+x_{j'}e_{j'}+x_{k'}e_{k'}+x_{l'}e_{l'})~~~~\text{(no summation)},\hspace{-1cm}
\ee
where $ijk\in\bold{L}$ is a line of the Fano plane and hence $i'j'k'l'\in\bold{Q}$ is the complementary quadrangle. An octonion $x$, written in this way, is a Kirmse integer if \{all of $x_0,x_i,x_j,x_k$ are integers \emph{or} all of $x_0,x_i,x_j,x_k$ are half-integers\} \emph{and} \{all of $x_{i'},x_{j'},x_{k'},x_{l'}$ are integers \emph{or} all of $x_{i'},x_{j'},x_{k'},x_{l'}$ are half-integers\}. As well as being well-packed, the Kirmse integers form the densest possible lattice in 8 dimensions. If around each lattice point we insert into $\R^8$ a ball with radius $\tfrac{1}{2}$, then every ball touches 240 others. In fact, this is none other than the root lattice of the largest exceptional group $E_8$.

Of particular importance to this paper are the 240 unit Kirmse integers -- elements of $\mathds{K}$ with unit norm -- given by:
%\be
%\left\{x\in\mathds{K}~\big|~|x|^2=1\right\}=\left\{\pm e_i,~\tfrac{1}{2}(\pm1\pm{e_i}\pm{e_j}\pm{e_k})~\text{with}~ijk\in\bold{L},~\tfrac{1}{2}(\pm e_i\pm e_j\pm e_k\pm e_l) ~\text{with} %~ijkl\in\bold{Q}\right\}.
%\ee
\be
\begin{split}
&\pm{1},\hspace{0.2cm}\pm{e_i},\\
&\tfrac{1}{2}(\pm1\pm{e_i}\pm{e_j}\pm{e_k})\hspace{0.2cm}\text{with}\hspace{0.2cm}ijk\in\bold{L},\\
&\tfrac{1}{2}(\pm{e_i}\pm{e_j}\pm{e_k}\pm{e_l})\hspace{0.2cm}\text{with}\hspace{0.2cm}ijkl\in\bold{Q}.
\end{split}
\ee
The set of Kirmse integers orthogonal to any particular basis element $e_a$ forms a copy of the root lattice of $E_7$. In particular, the set orthogonal to $e_0=1$ is just the pure-imaginary Kirmse integers, whose 126 unit elements are
\be
\pm{e}_i\hspace{0.2cm}\text{and}\hspace{0.2cm}\tfrac{1}{2}(\pm{e_i}\pm{e_j}\pm{e_k}\pm{e_l})\hspace{0.2cm}\text{with}\hspace{0.2cm}ijkl\in\bold{Q}.
\ee

In our discussion of the Kirmse integers we have so far overlooked the vital question of whether or not they are closed under multiplication. Kirmse himself once stated that they were. However, it is easy to find a counter-example:
\be
\tfrac{1}{2}(1+e_1+e_2+e_4)
\tfrac{1}{2}(1+e_2+e_3+e_5)=\tfrac{1}{2}(e_2+e_4+e_5+e_7)~\notin~\mathds{K}.
\ee
Hence $\mathds{K}$ is not closed under octonionic multiplication, a result sometimes referred to as \emph{Kirmse's mistake} \cite{Conway}. The mistake can be rectified by the following unusual trick. For every Kirmse integer $x=x_ae_a\in\mathds{K}$ exchange the coefficient $x_0$ with any one of the seven $x_i$. The resulting lattice $\mathds{K}'$ is just a reflection of the Kirmse lattice, and so is well-packed. However, in this case it \emph{is} closed under multiplication. In the literature, $\mathds{K}'$ has been referred to as the set of \emph{octavian integers} or the \emph{integral Cayley numbers}. We will demonstrate how they might be used in maximal supergravity theories in Section \ref{OCTAVIAN}.

\section{Octonionic Spinors in $D=11$}

A Majorana spinor in $D=11$ has 32 components, usually represented as a real column vector. Alternatively, viewing $\R^{32}$ as a tensor product $\R^4\otimes\R^8\cong\R^4\otimes\Oct\cong\Oct^4$, we can write the spinor as a 4-component octonionic column vector
\be\label{SPINOR}
\lambda = \begin{pmatrix} \lambda_1 \\ \lambda_2 \\ \lambda _3 \\ \lambda_4 \end{pmatrix},~~~~ \lambda_\alpha \in \Oct, ~~\alpha =1,2,3,4.
\ee
To generate the $4\times4$ octonionic Clifford algebra we seek a set of matrices $\{(\gamma^M)_\alpha{}^\beta\}$, $M=0,1,\ldots,10$, satisfying
\be
\gamma^M(\gamma^N\lambda)+\gamma^N(\gamma^M\lambda)=2\eta^{MN}\lambda~~~\forall~~~\lambda\in\Oct^4.
\ee
A natural choice \cite{Anastasiou:2014zfa} for these is the set
\be\label{GAMMAS}
\begin{split}
\gamma^0&=\begin{pmatrix}0&0 & 1&0\\ 0&0 & 0& 1\\-1&0 & 0&0\\0&-1 & 0&0\end{pmatrix},~~~
\gamma^{a+1}=\begin{pmatrix}0&0 & 0&e_a^*\\ 0&0 & e_a& 0\\0&e_a^* & 0&0\\e_a&0 & 0&0\end{pmatrix},\\
\gamma^9&=\begin{pmatrix}0&0 & 1&0\\ 0&0 & 0& -1\\1&0 & 0&0\\0&-1 & 0&0\end{pmatrix},~~~~
\gamma^{10}~=\begin{pmatrix}1~&0 & 0&0\\ 0~&1 & 0& 0\\0~&0 & -1&0\\0~&0 & 0&-1\end{pmatrix},
\end{split}
\ee
with $a=0,1,\ldots,7$. To see how these are related to a more familiar real $32\times 32$ set we can simply take their `matrix elements':
\be\label{MATRIXELEMENTS}
\begin{split}
\langle{e_a}|(\g^\mu)_\alpha{}^\beta{e_b}\rangle&=(\g^\mu)_\alpha{}^\beta\langle{e_a}|{e_b}\rangle=(\g^\mu)_\alpha{}^\beta\delta_{ab},~~~~~\mu=0,1,9,10, \\
\langle{e_a}|(\g^{i+1})_\alpha{}^\beta{e_b}\rangle&=(\g_*)_\alpha{}^\beta\langle{e_a}|e_i{e_b}\rangle=(\g_*)_\alpha{}^\beta\Gamma^i_{ab}~~~~~i=1,\cdots,7,
\end{split}
\ee
where $\g_*$ is defined by
\be
\g_*=-\g^0\g^1\g^9\g^{10},
\ee
and $\Gamma^i_{ab}$ is an element of the SO(7) Clifford algebra, formed from the octonionic structure constants:
\be
\Gamma^i_{ab}=\delta_{a0}\delta_{ib}-\delta_{b0}\delta_{ia}+C_{iab}.
\ee
We see that writing the gamma matrices over the octonions corresponds to an $\bold{11=4+7}$ split:
\be
\SO(1,10)\supset \SO(1,3)\times\SO(7),
\ee
with the seven imaginary octonions playing the role of the SO(7) gamma matrices and the four real $\g^\mu$, $\mu=0,1,9,10$, playing the role of the (`really real' Majorana) SO(1,3) gamma matrices. An obvious appeal of this octonionic parameterisation is that this natural split associates the seven extra dimensions of M-theory with the seven imaginary octonionic basis elements. 

By equation \eqref{MATRIXELEMENTS}, left-multiplying $\lambda\in\Oct^4$ by the octonionic matrix $\g^M$ corresponds to multiplying $\lambda$'s 32 real components by an ordinary real $32\times 32$ gamma matrix. By successive composition we deduce that in general, the action of the rank $r$ Clifford algebra element on $\lambda$ can be written
\be\label{rankr}
\g^{[M_1}\Big(\g^{M_2}\big(\ldots(\g^{M_{r-1}}(\g^{M_r]}\lambda))\ldots\big)\Big).
\ee
The positioning of the brackets fixes any ambiguities due to non-associativity. For example, an infinitesimal Lorentz transformation of a spinor $\lambda$ is
\be
\delta\lambda=\frac{1}{4}\omega_{MN}\gamma^M(\gamma^N\lambda),
\ee
where $\omega_{MN}=-\omega_{NM}$. 

As in \cite{Anastasiou:2014zfa}, let us define an operator $\ghat^M$, whose action is left-multiplication by $\g^M$, so that we can think of the rank $r$ Clifford algebra element as the operator
\be
\ghat^{M_1M_2\ldots M_r}\equiv\ghat^{[M_1}\ghat^{M_2}\ldots\ghat^{M_r]},
\ee
where the operators $\ghat^M$ must be composed as
\be
\ghat^M\ghat^N\lambda=\g^M(\g^N\lambda)\neq (\g^M\g^N)\lambda.
\ee
This ensures that the action of $\ghat^{[M_1M_2\ldots M_r]}$ on a spinor is given by (\ref{rankr}), as required.

In order to construct the supergravity Lagrangian and transformation rules in this language, we will require real spinor bilinears. These are built using the charge conjugation matrix $C^{\alpha\beta}$ (which is numerically equal to $\g^0$ but with a different index structure)
\be
C^{\alpha\beta}=\begin{pmatrix}0&0 & 1&0\\ 0&0 & 0& 1\\-1&0 & 0&0\\0&-1 & 0&0\end{pmatrix}.
\ee
This matrix satisfies
\be
C(\g^M)^\dagger C=\g^M ~~~\Leftrightarrow~~~ (\g^M)^\dagger C=-C\g^M,
\ee
where the dagger denotes transposition and octonionic conjugation. Let us define
\be
\bar\lambda:=\lambda^\dagger C.
\ee
If $\lambda_1$ and $\lambda_2$ are octonionic spinors whose real components are anti-commuting Grassmann numbers, then the quantity 
\be
\text{Re}(i\bar\lambda_1\lambda_2)=\frac{1}{2}\left(i\bar\lambda_1\lambda_2+(i\bar\lambda_1\lambda_2)^\dagger\right)=\frac{i}{2}\left(\bar\lambda_1\lambda_2+\bar\lambda_2\lambda_1\right)
\ee
is Lorentz-invariant. Note that we define the dagger operation such that it also complex-conjugates the factor of $i$. This accounts for the anti-commuting spinor components. We can now form a general spinor bilinear as follows:
\be
\text{Re}(i\bar\lambda_1\ghat^{M_1M_2\ldots M_r}\lambda_2),
\ee
which will then transform as an $r$-index antisymmetric tensor under Lorentz transformations.

\section{The Lagrangian and Transformation Rules}

With the tools described above it is not difficult to rewrite the Lagrangian and transformation rules of $D=11$ supergravity over the octonions. Starting from the conventional Lagrangian, all one must do is exchange any 32-component real spinors with their 4-component octonionic counterparts, and exchange any bilinears with those described above. This gives the following Lagrangian:
\begin{align}\label{LAGRANGIAN}
\mathcal{L}=\sqrt{-g}\Bigg[&R-\text{Re}\left(i\bar\Psi_M\hat\gamma^{MNP}D_N\left(\frac{1}{2}(\omega+\tilde\omega)\right)\Psi_P\right)-\frac{1}{24}F_{MNPQ}F^{MNPQ}\\
 &-\frac{\sqrt{2}}{192}\nonumber\text{Re}\left(i\bar\Psi_R\left(\hat\gamma^{MNPQRS}+12\hat\gamma^{MN}g^{PR}g^{QS}\right)\Psi_S\right)
\left(F_{MNPQ}+\tilde{F}_{MNPQ}\right)\\ \nonumber
&-\frac{2\sqrt{2}}{144^2}\varepsilon^{M_0M_1\cdots M_{10}}F_{M_0M_1M_2M_3}F_{M_4M_5M_6M_7}A_{M_8M_9M_{10}} \Bigg], \nonumber
\end{align}
where
\be
\begin{split}
\omega_M{}^{AB}&=\omega_M{}^{AB}(e)+K_M{}^{AB},\\
\tilde\omega_M{}^{AB}&=\omega_M{}^{AB}(e)-\frac{1}{4}\text{Re}\left(i(\bar\Psi_M\hat\g^B\Psi^A-\bar\Psi^A\hat\g_M\Psi^B+\bar\Psi^B\hat\g^A\Psi_M)\right),\\
K_M{}^{AB}&=-\frac{1}{4}\text{Re}\left(i(\bar\Psi_M\hat\g^B\Psi^A-\bar\Psi^A\hat\g_M\Psi^B+\bar\Psi^B\hat\g^A\Psi_M)\right),\\
\tilde{F}_{MNPQ}&=4\partial_{[M}A_{NPQ]}+\frac{3\sqrt{2}}{2}\text{Re}\left(i\bar\Psi_{[M}\hat\g_{NP}\Psi_{Q]}\right),
\end{split}
\ee
and the covariant derivative ${D}_M(\omega)$ is defined by
\be
{D}_M(\omega)\epsilon=\partial_M\epsilon+\frac{1}{4}\omega_M{}^{AB}\hat\g_{AB}\epsilon.
\ee
The Lagrangian \eqref{LAGRANGIAN} is invariant under the following supersymmetry transformations:
\be
\begin{split}
\delta e_M^A&=\frac{1}{2}\text{Re}\left(i\bar\epsilon\hat\gamma^{A}\Psi_{M}\right),\\ 
\delta C_{MNP}&=-\frac{3\sqrt{2}}{4}\text{Re}\left(i\bar\epsilon\hat\gamma_{[MN}\Psi_{P]}\right),\\
\delta\Psi_M&={D}_M(\tilde\omega)\epsilon+\frac{\sqrt{2}}{288}(\hat\g^{ABCD}{}_M-8\delta^A_M\hat\g^{BCD})\tilde{F}_{ABCD}\epsilon,
\end{split}
\ee
although we do not prove this here.

\section{The Octavian Integers as Dilaton Vectors}\label{OCTAVIAN}

\subsection{$D=4$, $\N=8$ Supergravity}

Next we consider dimensional reduction to $D=4$, yielding $\N=8$ supergravity. Dropping the dependence of the fields on the seven coordinates associated with the seven imaginary basis octonions (as in equation \eqref{GAMMAS}), we find the following bosonic content:
\be
\begin{split}
g_{MN} ~~&\rightarrow~~ g_{\mu\nu},~\vec{\phi},~\mathcal{A}^i_\mu,~\mathcal{A}^i{}_j ~(\text{with}~i<j),\\
A_{MNP}~~&\rightarrow~~ A_{\mu\nu\rho},~A_{\mu\nu i},~A_{\mu ij},~A_{ijk},
\end{split}
\ee
where $i,j,k$ run over the seven internal dimensions, while $\mu,\nu,\rho$ run over the extended four, and $\vec{\phi}$ denotes the seven dilatons written as a seven-component vector. Note that the scalar fields descended from $g_{MN}$ have not been written here so as to be covariant with respect to the SO(7) (or GL$(7,\R)$) symmetry associated with the internal dimensions; instead we have separated the dilatonic and axionic scalars as $\vec{\phi}$ and $\mathcal{A}^i{}_j$. 

Denoting $(p+1)$-form field strengths of $p$-form potentials with superscripts $(p+1)$, the Lagrangian for the bosonic sector is then:
\be
\begin{split}\label{N=8LAGRANGIAN}
\mathcal{L}_{B}=~\sqrt{-g}\Big[&R-\frac{1}{2}(\partial\vec{\phi})^2
-\frac{1}{2}\sum_{i}e^{2\vec{b}_{i}\cdot\vec{\phi}}(\mathcal{F}^{(2)}_{i})^2
-\frac{1}{2}\sum_{i<j}e^{2\vec{b}_{ij}\cdot\vec{\phi}}(\mathcal{F}^{(1)}_{ij})^2
-\frac{1}{2}e^{2\vec{a}\cdot\vec{\phi}}({F}^{(4)})^2
\\
&-\frac{1}{2}\sum_{i}e^{2\vec{a}_{i}\cdot\vec{\phi}}({F}^{(3)}_{i})^2
-\frac{1}{2}\sum_{i<j}e^{2\vec{a}_{ij}\cdot\vec{\phi}}({F}^{(2)}_{ij})^2
-\frac{1}{2}\sum_{i<j<k}e^{2\vec{a}_{ijk}\cdot\vec{\phi}}({F}^{(1)}_{ijk})^2
+\mathcal{L}_{FFA}\Big],
\end{split}
\ee
where the field strengths and their transgression terms are defined in \cite{Lu:1995yn,PhysRevD.83.046007} and $\mathcal{L}_{FFA}$ denotes the terms descended from the topological term in the eleven dimensional Lagrangian. Note that in the bosonic sector all the fields are real, since the octonions have so far only been used in the description of the fermions. The constant `dilaton vectors' $\vec{a}$, $\vec{a}_{i}$, $\vec{a}_{ij}$, $\vec{a}_{ijk}$, $\vec{b}_{i}$ and $\vec{b}_{ij}$ parameterise the non-canonical coupling of the seven dilatons $\vec{\phi}$ to the other bosonic fields. For the various potentials, they are given by:
\begin{align}
&\text{3-potential:}&&\vec{a}=-\vec{g},&&&&\nonumber\\ 
&\text{2-potentials:}&&\vec{a}_{i}=\vec{f_i}-\vec{g},\nonumber\\
&\text{1-potentials:}&&\vec{a}_{ij}=\vec{f_i}+\vec{f_j}-\vec{g},&&\vec{b}_{i}=-\vec{f_i},\nonumber\\
&\text{0-potentials:}&&\vec{a}_{ijk}=\vec{f_i}+\vec{f_j}+\vec{f_k}-\vec{g},&&\vec{b}_{ij}=-\vec{f_i}+\vec{f_j},\label{DVECTORS}
\end{align}
where the vectors $\vec{g}$ and $\vec{f_i}$ satisfy
\begin{align}\label{SCALARPRODUCTS}
\vec{g}=\tfrac{1}{3}\sum_{i}\vec{f_i},&&\vec{g}\cdot\vec{g}=\frac{11-D}{2(D-2)},&&\vec{g}\cdot\vec{f}_i=\frac{3}{2(D-2)},&&\vec{f}_{i}\cdot\vec{f}_{j}=\frac{\delta_{ij}}{2}+\frac{1}{2(D-2)}.&&
\end{align}
Traditionally these vectors are given as below \cite{Lu:1995yn} -- presented for $D$ extended dimensions (temporarily letting $i,j =1,2\cdots,(11-D)$ for these few equations) so that we can use the expressions later on for $D=3$:
\begin{align}
&\vec{f}_{i}=\tfrac{1}{2}(\underbrace{0,\dots,0}_{i-1},(10-i)s^i,s^{i+1},\dots{s}^{11-D})\nonumber\\
&\vec{g}=\tfrac{3}{2}(s^1,s^2,\dots,s^{11-D})\\
&s^i=\sqrt{2/((10-i)(9-i))}.\nonumber
\end{align}
However, in the following we present our own alternative octonionic parameterisation, which makes manifest the relationship between the bosonic sector, the Fano plane and the dual Fano plane.

Returning to $D=4$, we use $\vec{g}$ and the seven $\vec{f}_i$ to compute all the dilaton vectors. In particular, $\vec{a}_{ijk}$, $\vec{b}_{ij}$ and $-\vec{a}_i$, the vectors parameterising the coupling of the dilatons to the 63 axions, are the positive roots of the U-duality group $E_{7(7)}$ (where we dualise the seven 2-forms $A_{\mu\nu i}$ to scalars, whose dilaton vectors are $-\vec{a}_i$). The dilaton vectors $\vec{a}_{ij}$ and $\vec{b}_i$ make up the positive weights of the $\bold{56}$ of $E_{7(7)}$, under which the 2-form field strengths and their duals transform.

Since we have a dilaton for each internal dimension, and the seven internal dimensions are associated with the seven imaginary octonions via \eqref{GAMMAS}, it makes sense to consider the seven dilatons $\vec\phi$ themselves to be components of an imaginary octonion $\phi_ie_i$. In this case, the dilaton vectors should also be viewed as a particular set of imaginary octonions, so that we may form the scalar products that appear in the exponential couplings in \eqref{LAGRANGIAN}. We will demonstrate that this perspective has some interesting consequences.

Consider the replacement $\vec{f}_i \rightarrow f_i\in\text{Im}\,\Oct$, where
\begin{align}
&&f_1=\tfrac{1}{2}(e_1+e_2+e_4),&&f_2=\tfrac{1}{2}(e_2+e_3+e_5),&&f_3=\tfrac{1}{2}(e_3+e_4+e_6),&&f_4=\tfrac{1}{2}(e_4+e_5+e_7),\nonumber\\&&f_5=\tfrac{1}{2}(e_5+e_6+e_1),&&f_6=\tfrac{1}{2}(e_6+e_7+e_2),&&f_7=\tfrac{1}{2}(e_7+e_1+e_3),&& \label{D4F}
\end{align}
i.e. $f_i=\tfrac{1}{2}(e_i+e_j+e_k)$, with $ijk\in\bold{L}$. This amounts only to a change of basis in the space of dilaton vectors. As clarified below, this particular choice of parameterisation makes manifest the relationship between the bosonic fields of $\N=8$ supergravity, the Fano plane and the dual Fano plane. We find that
\be
g=\tfrac{1}{3}\sum_{i}{f_i}=\tfrac{1}{2}(e_1+e_2+e_3+e_4+e_5+e_6+e_7),
\ee
and using the inner product defined in \eqref{INNER}, one can check that
\begin{align}
\langle{g}|{g}\rangle=\frac{7}{4},&&\langle{g}|{f_i}\rangle=\frac{3}{4},&&\langle{f_i}|{f_j}\rangle=\frac{\delta_{ij}}{2}+\frac{1}{4},
\end{align}
as required when $D=4$.

\begin{figure}[h!]
  \centering
    \includegraphics[width=0.65\textwidth]{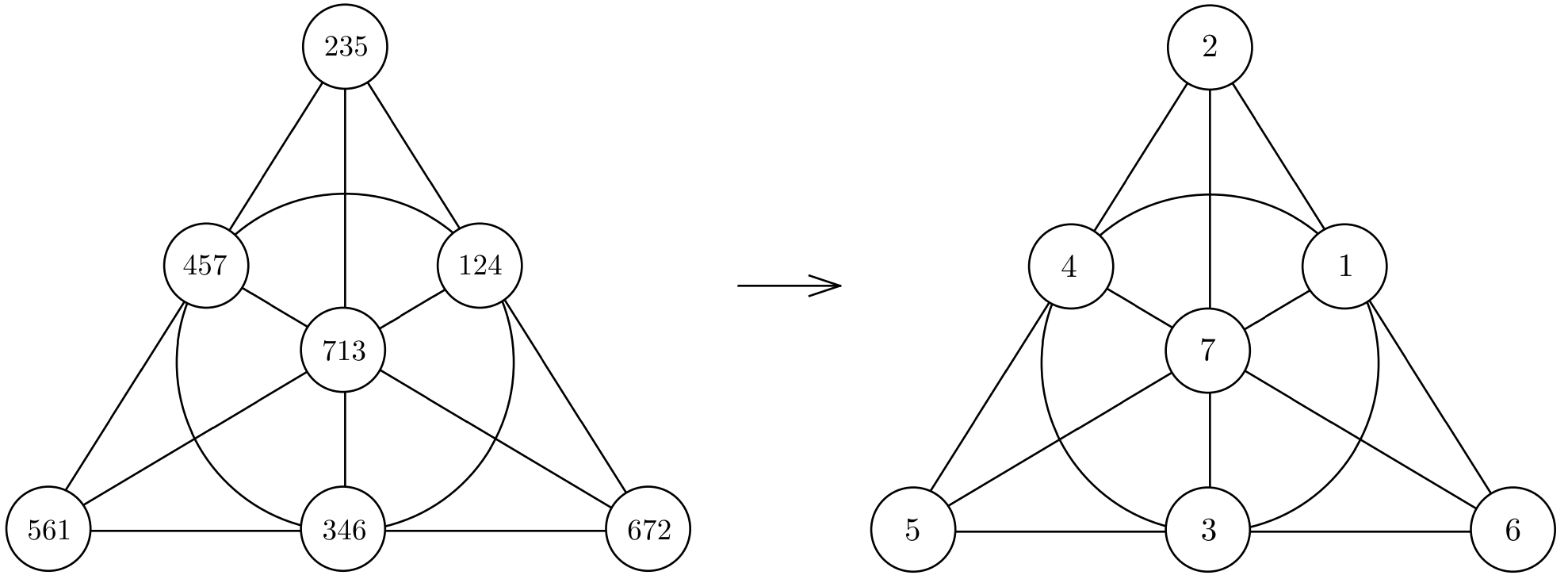}
  \caption{\footnotesize{The dual Fano plane $\widetilde{\F}$ obtained by interchanging the roles of points and lines on the original Fano plane. Relabelling the triples 124, 235, 346, 457, 561, 672, 713 $\rightarrow$ 1, 2, 3, 4, 5, 6, 7 gives the plane on the right. We have not chosen orientations for the lines of the dual Fano plane since we do not use it for multiplication.}}\label{DUALFANO}
\end{figure}

Now, using \eqref{DVECTORS} we can compute the various dilaton vectors in this octonionic parameterisation. However, before we do so, it will be useful to properly introduce the dual Fano plane. In general, a projective plane $\mathds{P}$ exhibits a duality between its points and lines, whose roles may be interchanged to obtain a new space $\widetilde{\mathds{P}}\cong\mathds{P}$. For every statement relating points and lines on $\mathds{P}$ there is a dual statement relating \emph{lines and points} on $\widetilde{\mathds{P}}$. For example, just as two points on a projective plane lie on a unique line, two lines on the plane meet at a unique point. Since the (unoriented) Fano plane is the projective plane over the field $\mathds{Z}_2$, if we interchange the roles of its points and lines we obtain a dual plane - see Fig. \ref{DUALFANO}. 

In practice we relabel the lines 124, 235, 346, 457, 561, 672, 713 as 1, 2, 3, 4, 5, 6, 7, respectively, which leads to the plane on the right in Fig. \ref{DUALFANO}, whose lines are given by the set $\widetilde{\bold{L}}=\{157,261,372,413,524,635,746\}$. This relabelling is deliberately chosen so as to match up with \eqref{D4F}.

Now we compute the $E_{7(7)}$ root dilaton vectors, starting with those whose expressions are simplest. Since a line in $\widetilde{\bold{L}}$ corresponds to a point in the original Fano plane $\mathds{F}$, we should expect $a_{ijk}$ with $ijk\in\widetilde{\bold{L}}$ to correspond in some way to a point in $\mathds{F}$. This is indeed the case, since
\begin{align}
a_{157}=e_1,&&a_{261}=e_2,&&a_{372}=e_3,&&a_{413}=e_4,&&a_{524}=e_5,&&a_{635}=e_6,&&a_{746}=e_7.&&
\end{align}
Next we consider the octonions $-a_i$, whose labels correspond to points on the dual Fano plane $\widetilde{\mathds{F}}$ and hence to lines on $\mathds{F}$. Indeed, one finds that
\begin{align}
&&-a_1\hspace{-0.04cm}=\hspace{-0.04cm}\tfrac{1}{2}(e_3\hspace{-0.07cm}+\hspace{-0.07cm}e_5\hspace{-0.07cm}+\hspace{-0.07cm}e_6\hspace{-0.07cm}+\hspace{-0.07cm}e_7),&&-a_2\hspace{-0.04cm}=\hspace{-0.04cm}\tfrac{1}{2}(e_4\hspace{-0.07cm}+\hspace{-0.07cm}e_6\hspace{-0.07cm}+\hspace{-0.07cm}e_7\hspace{-0.07cm}+\hspace{-0.07cm}e_1),&&-a_3\hspace{-0.04cm}=\hspace{-0.04cm}\tfrac{1}{2}(e_5\hspace{-0.07cm}+\hspace{-0.07cm}e_7\hspace{-0.07cm}+\hspace{-0.07cm}e_1\hspace{-0.07cm}+\hspace{-0.07cm}e_2),&&-a_4\hspace{-0.04cm}=\hspace{-0.04cm}\tfrac{1}{2}(e_6\hspace{-0.07cm}+\hspace{-0.07cm}e_1\hspace{-0.07cm}+\hspace{-0.07cm}e_2\hspace{-0.07cm}+\hspace{-0.07cm}e_3),\nonumber\\
&&-a_5\hspace{-0.04cm}=\hspace{-0.04cm}\tfrac{1}{2}(e_7\hspace{-0.07cm}+\hspace{-0.07cm}e_2\hspace{-0.07cm}+\hspace{-0.07cm}e_3\hspace{-0.07cm}+\hspace{-0.07cm}e_4),&&-a_6\hspace{-0.04cm}=\hspace{-0.04cm}\tfrac{1}{2}(e_1\hspace{-0.07cm}+\hspace{-0.07cm}e_3\hspace{-0.07cm}+\hspace{-0.07cm}e_4\hspace{-0.07cm}+\hspace{-0.07cm}e_5),&&-a_7\hspace{-0.04cm}=\hspace{-0.04cm}\tfrac{1}{2}(e_2\hspace{-0.07cm}+\hspace{-0.07cm}e_4\hspace{-0.07cm}+\hspace{-0.07cm}e_5\hspace{-0.07cm}+\hspace{-0.07cm}e_6),&&
\end{align}
which match up with the seven quadrangles complimentary to the seven corresponding lines of $\mathds{F}$. Computing the rest of the vectors, $a_{ijk}$ ($ijk\notin\widetilde{\bold{L}}$) and $b_{ij}$ (see Appendix), we find that the whole set populates the unit imaginary Kirmse integers:
\be
\pm{e}_i\hspace{0.2cm}\text{and}\hspace{0.2cm}\tfrac{1}{2}(\pm{e_i}\pm{e_j}\pm{e_k}\pm{e_l})\hspace{0.2cm}\text{with}\hspace{0.2cm}ijkl\in\bold{Q}.
\ee
We will show in Section \ref{FOURCURIOUS} how this parameterisation inspires, from a new perspective, a series of natural truncations that has appeared before in the literature. The vectors $a_{ij}$ and $b_i$ corresponding to the 1-form gauge potentials all have the form
\be
\tfrac{1}{2}(\pm{e_i}\pm{e_j}\pm{e_k})\hspace{0.2cm}\text{with}\hspace{0.2cm}ijkl\in\bold{L}.
\ee
Putting all of this together means that (after dualisation) we can schematically write the bosonic $\N=8$ Lagrangian as 
\be
\mathcal{L}={\sqrt{-g}} \Bigg[
R
-\frac{1}{2}  \langle\partial\phi|\partial\phi\rangle
-\frac{1}{2} \sum_{\text{\scriptsize{points}}}e^{\langle\text{\scriptsize{points}}|\phi\rangle} (F_{\text{\scriptsize{points}}}^{(1)})^2
-\frac{1}{2} \sum_{\text{\scriptsize{quads}}}e^{\langle\text{\scriptsize{quads}}|\phi\rangle} ({ F}_{\text{\scriptsize{quads}}}^{(1)})^2
-\frac{1}{4} \sum_{\text{\scriptsize{lines}}} e^{\langle\text{\scriptsize{lines}}|\phi\rangle} ({ F}_{\text{\scriptsize{lines}}}^{(2)})^2\Bigg],
\ee
where the sums run over all the vectors listed in Table \ref{D=4FULL} in the Appendix, which correspond to the points, lines and quadrangles of the Fano plane with the various possible $\pm$ sign combinations. As alluded to above, this new parameterisation makes manifest the relationship between the bosonic fields and the structure of the Fano plane.

Before we move on to $D=3$, we will briefly demonstrate a nice Fano-plane-based trick for restricting the roots of $E_{7(7)}$ to those of its maximal compact subgroup $\SU(8)$. The adjoint of $E_{7(7)}$ decomposes into SU(8) as:
\be
\bold{133}\rightarrow\bold{63+70},
\ee
so we expect the 126 roots of  $E_{7(7)}$ to split into two sets: a set consisting of the 56 roots of SU(8) and the remaining 70 vectors corresponding to the weights of the $\bold{70}$ representation. The trick is first to choose a line of the Fano plane -- let us say 124. As shown in \cite{Koca2007808}, we then discard the unit Kirmse integers $\pm e_1$, $\pm e_2$ and $\pm e_4$, as well as those associated with the corresponding quadrangle -- in this case $\tfrac{1}{2}(\pm e_3\hspace{-0.07cm}\pm\hspace{-0.07cm}e_5\hspace{-0.07cm}\pm\hspace{-0.07cm}e_6\hspace{-0.07cm}\pm\hspace{-0.07cm}e_7)$. We then take the remaining quadrangles and wherever $e_1$, $e_2$ and $e_4$ appear we fix their relative signs according to the following rule: if we choose another point -- say $e_7$ -- then the signs are the same if $e_7$ appears in the quadrangle and different if it does not. This is shown explicitly in Table \ref{SU8}.

\begin{table}[h!]
$\begin{array}{c|cccccccc}
\toprule
&\\
E_7 ~\text{roots}&&SU(8)~\text{roots}\\
&\\
\hline
&\\
\pm e_1, \pm e_2, \pm e_3, \pm e_4, \pm e_5,  \pm e_6, \pm e_7&&\pm e_3,
\pm e_5, \pm e_6,  \pm e_7\\
 \frac{1}{2}( \pm e_3 \pm  e_5 \pm e_6 \pm e_7)&& \\
 \frac{1}{2}( \pm e_4 \pm  e_6 \pm e_7 \pm e_1)&& \pm\frac{1}{2}( +
e_4 \pm  e_6 \pm e_7 + e_1)\\
 \frac{1}{2}( \pm e_5 \pm  e_7 \pm e_1 \pm e_2)&&\pm\frac{1}{2}( \pm e_5
\pm  e_7 + e_1 + e_2)\\
 \frac{1}{2}( \pm e_6 \pm  e_1 \pm e_2 \pm e_3)&& \pm\frac{1}{2}( \pm e_6
+  e_1 - e_2 \pm e_3)\\
 \frac{1}{2}( \pm e_7 \pm  e_2 \pm e_3 \pm e_4)&& \pm\frac{1}{2}( \pm e_7
+  e_2 \pm e_3 + e_4)\\
 \frac{1}{2}( \pm e_1\pm  e_ 3\pm e_4 \pm e_5)&& \pm\frac{1}{2}( +
e_1\pm  e_ 3 - e_4 \pm e_5)\\
 \frac{1}{2}( \pm e_2 \pm  e_4 \pm e_5 \pm e_6)&& \pm\frac{1}{2}(  +
e_2 -  e_4 \pm e_5 \pm e_6)\\
 &\\
 \hline
\hline
\end{array}$
\caption{$E_7\supset SU(8)$ roots}\label{e7}\label{SU8}
\end{table}

\subsection{$D=3$ , $\mathcal{N}=16$ Supergravity}

Dimensionally reducing to $D=3$ means there are eight internal dimensions, which we label $a=0,1,\dots,7$. As a result, there are eight dilatons $\vec{\phi}$, which we now write as an octonion $\phi_a e_a$. Accordingly, we replace the eight vectors $\vec{f_a}$, each of which has eight entries, with eight octonions: $\vec{f}_a \rightarrow f_a\in\Oct$, where we choose the parameterisation
\begin{align}
&&f_0=1,&&f_1=\tfrac{1}{2}(1\hspace{-0.07cm}+\hspace{-0.07cm}e_1\hspace{-0.07cm}+\hspace{-0.07cm}e_2\hspace{-0.07cm}+\hspace{-0.07cm}e_4),
&&f_2=\tfrac{1}{2}(1\hspace{-0.07cm}+\hspace{-0.07cm}e_2\hspace{-0.07cm}+\hspace{-0.07cm}e_3\hspace{-0.07cm}+\hspace{-0.07cm}e_5),
&&f_3=\tfrac{1}{2}(1\hspace{-0.07cm}+\hspace{-0.07cm}e_3\hspace{-0.07cm}+\hspace{-0.07cm}e_4\hspace{-0.07cm}+\hspace{-0.07cm}e_6),\nonumber\\
&&f_5=\tfrac{1}{2}(1\hspace{-0.07cm}+\hspace{-0.07cm}e_5\hspace{-0.07cm}+\hspace{-0.07cm}e_6\hspace{-0.07cm}+\hspace{-0.07cm}e_1),
&&f_6=\tfrac{1}{2}(1\hspace{-0.07cm}+\hspace{-0.07cm}e_6\hspace{-0.07cm}+\hspace{-0.07cm}e_7\hspace{-0.07cm}+\hspace{-0.07cm}e_2),
&&f_7=\tfrac{1}{2}(1\hspace{-0.07cm}+\hspace{-0.07cm}e_7\hspace{-0.07cm}+\hspace{-0.07cm}e_1\hspace{-0.07cm}+\hspace{-0.07cm}e_3),
&&f_4=\tfrac{1}{2}(1\hspace{-0.07cm}+\hspace{-0.07cm}e_4\hspace{-0.07cm}+\hspace{-0.07cm}e_5\hspace{-0.07cm}+\hspace{-0.07cm}e_7).&& \label{D3F}
\end{align}
This gives
\be
g=\tfrac{1}{2}(3+e_1+e_2+e_3+e_4+e_5+e_6+e_7),
\ee
so that calculating the inner products we find
\begin{align}
\langle{g}|{g}\rangle=4,&&\langle{g}|{f_a}\rangle=\frac{3}{2},&&\langle{f_a}|{f_b}\rangle=\frac{\delta_{ab}}{2}
+\frac{1}{2},
\end{align}
consistent with \eqref{SCALARPRODUCTS} for $D=3$.

In $D=3$ the 2-form potentials carry no degrees of freedom and the 1-form potentials may be dualised to scalars. Since the metric contains no dynamical degrees of freedom, all the bosonic degrees of freedom of the theory are carried by the resulting 128 scalars, whose dilaton vectors are $-\vec{a}_{ab}$, $\vec{a}_{abc}$, $-\vec{b}_{a}$ and $\vec{b}_{ab}$. These make up the positive roots of the U-duality group of the theory, $E_{8(8)}$.

Again, due to the labelling system chosen in \eqref{D3F}, we should expect the $a_{abc}$ with $abc\in\widetilde{\bold{L}}$ to correspond in some way to the points of the Fano plane $\F$. This is indeed the case:
\begin{align}
a_{157}=e_1,&&a_{261}=e_2,&&a_{372}=e_3,&&a_{413}=e_4,&&a_{524}=e_5,&&a_{635}=e_6,&&a_{746}=e_7.
\end{align}
Similarly, the $b_a$ correspond to points on $\widetilde\F$ and hence to lines on $\F$:
\begin{align}
&&-b_0=1,&&-b_1=\tfrac{1}{2}(1\hspace{-0.07cm}+\hspace{-0.07cm}e_1\hspace{-0.07cm}+\hspace{-0.07cm}e_2\hspace{-0.07cm}+\hspace{-0.07cm}e_4),
&&-b_2=\tfrac{1}{2}(1\hspace{-0.07cm}+\hspace{-0.07cm}e_2\hspace{-0.07cm}+\hspace{-0.07cm}e_3\hspace{-0.07cm}+\hspace{-0.07cm}e_5),
&&-b_3=\tfrac{1}{2}(1\hspace{-0.07cm}+\hspace{-0.07cm}e_3\hspace{-0.07cm}+\hspace{-0.07cm}e_4\hspace{-0.07cm}+\hspace{-0.07cm}e_6),\nonumber\\
&&-b_4=\tfrac{1}{2}(1\hspace{-0.07cm}+\hspace{-0.07cm}e_4\hspace{-0.07cm}+\hspace{-0.07cm}e_5\hspace{-0.07cm}+\hspace{-0.07cm}e_7),
&&-b_5=\tfrac{1}{2}(1\hspace{-0.07cm}+\hspace{-0.07cm}e_5\hspace{-0.07cm}+\hspace{-0.07cm}e_6\hspace{-0.07cm}+\hspace{-0.07cm}e_1),
&&-b_6=\tfrac{1}{2}(1\hspace{-0.07cm}+\hspace{-0.07cm}e_6\hspace{-0.07cm}+\hspace{-0.07cm}e_7\hspace{-0.07cm}+\hspace{-0.07cm}e_2),
&&-b_7=\tfrac{1}{2}(1\hspace{-0.07cm}+\hspace{-0.07cm}e_7\hspace{-0.07cm}+\hspace{-0.07cm}e_1\hspace{-0.07cm}+\hspace{-0.07cm}e_3).
&& 
\end{align}
The $a_{0i}$ also reflect this simple correspondence:
\begin{align}
&&-a_{01}\hspace{-0.04cm}=\hspace{-0.04cm}\tfrac{1}{2}(e_3\hspace{-0.07cm}+\hspace{-0.07cm}e_5\hspace{-0.07cm}+\hspace{-0.07cm}e_6\hspace{-0.07cm}+\hspace{-0.07cm}e_7), &&-a_{02}\hspace{-0.04cm}=\hspace{-0.04cm}\tfrac{1}{2}(e_4\hspace{-0.07cm}+\hspace{-0.07cm}e_6\hspace{-0.07cm}+\hspace{-0.07cm}e_7\hspace{-0.07cm}+\hspace{-0.07cm}e_1), &&-a_{03}\hspace{-0.04cm}=\hspace{-0.04cm}\tfrac{1}{2}(e_5\hspace{-0.07cm}+\hspace{-0.07cm}e_7\hspace{-0.07cm}+\hspace{-0.07cm}e_1\hspace{-0.07cm}+\hspace{-0.07cm}e_2), &&-a_{04}\hspace{-0.04cm}=\hspace{-0.04cm}\tfrac{1}{2}(e_6\hspace{-0.07cm}+\hspace{-0.07cm}e_1\hspace{-0.07cm}+\hspace{-0.07cm}e_2\hspace{-0.07cm}+\hspace{-0.07cm}e_3),\nonumber\\
&&-a_{05}\hspace{-0.04cm}=\hspace{-0.04cm}\tfrac{1}{2}(e_7\hspace{-0.07cm}+\hspace{-0.07cm}e_2\hspace{-0.07cm}+\hspace{-0.07cm}e_3\hspace{-0.07cm}+\hspace{-0.07cm}e_4), &&-a_{06}\hspace{-0.04cm}=\hspace{-0.04cm}\tfrac{1}{2}(e_1\hspace{-0.07cm}+\hspace{-0.07cm}e_3\hspace{-0.07cm}+\hspace{-0.07cm}e_4\hspace{-0.07cm}+\hspace{-0.07cm}e_5), &&-a_{07}\hspace{-0.04cm}=\hspace{-0.04cm}\tfrac{1}{2}(e_2\hspace{-0.07cm}+\hspace{-0.07cm}e_4\hspace{-0.07cm}+\hspace{-0.07cm}e_5\hspace{-0.07cm}+\hspace{-0.07cm}e_6),
\end{align}
as well as the $b_{0i}$:
\begin{align}
&&b_{01}=\tfrac{1}{2}(-1\hspace{-0.07cm}+\hspace{-0.07cm}e_1\hspace{-0.07cm}+\hspace{-0.07cm}e_2\hspace{-0.07cm}+\hspace{-0.07cm}e_4),
&&b_{02}=\tfrac{1}{2}(-1\hspace{-0.07cm}+\hspace{-0.07cm}e_2\hspace{-0.07cm}+\hspace{-0.07cm}e_3\hspace{-0.07cm}+\hspace{-0.07cm}e_5),
&&b_{03}=\tfrac{1}{2}(-1\hspace{-0.07cm}+\hspace{-0.07cm}e_3\hspace{-0.07cm}+\hspace{-0.07cm}e_4\hspace{-0.07cm}+\hspace{-0.07cm}e_6),
&&b_{04}=\tfrac{1}{2}(-1\hspace{-0.07cm}+\hspace{-0.07cm}e_4\hspace{-0.07cm}+\hspace{-0.07cm}e_5\hspace{-0.07cm}+\hspace{-0.07cm}e_7),
\nonumber\\
&&b_{05}=\tfrac{1}{2}(-1\hspace{-0.07cm}+\hspace{-0.07cm}e_5\hspace{-0.07cm}+\hspace{-0.07cm}e_6\hspace{-0.07cm}+\hspace{-0.07cm}e_1),
&&b_{06}=\tfrac{1}{2}(-1\hspace{-0.07cm}+\hspace{-0.07cm}e_6\hspace{-0.07cm}+\hspace{-0.07cm}e_7\hspace{-0.07cm}+\hspace{-0.07cm}e_2),
&&b_{07}=\tfrac{1}{2}(-1\hspace{-0.07cm}+\hspace{-0.07cm}e_7\hspace{-0.07cm}+\hspace{-0.07cm}e_1\hspace{-0.07cm}+\hspace{-0.07cm}e_3).
\end{align}
Computing all positive and negative roots (see Appendix), we recover the whole set of 240 unit Kirmse integers, the roots of $E_{8(8)}$:
\be
\begin{split}
&\pm{1},~~\pm{e_i},\\
&\tfrac{1}{2}(\pm1\pm{e_i}\pm{e_j}\pm{e_k})\hspace{0.2cm}\text{with}\hspace{0.2cm}ijk\in\bold{L},\\
&\tfrac{1}{2}(\pm{e_i}\pm{e_j}\pm{e_k}\pm{e_l})\hspace{0.2cm}\text{with}\hspace{0.2cm}ijkl\in\bold{Q}.
\end{split}
\ee

We have chosen to parameterise the vectors $\vec{f}_a$ as in equation \eqref{D3F} because this leads us to dilaton vectors that are easily recognisable as Kirmse integers (since the Kirmse integers take their structure from the lines and quadrangles of the Fano plane). However, we could just as easily have parameterised so as to arrive at the octavian integers, which are closed under multiplication. In other words, in the manor above, the dilaton vectors of $D=3,~\mathcal{N}=16$ supergravity may be equipped with a multiplication rule, under which they form a closed algebra.

Just as in $D=4$ above, in $D=3$ we can also write the dualised $\N=16$ bosonic Lagrangian as
\be
\mathcal{L}={\sqrt{-g}} \Bigg[
R
-\frac{1}{2}  \langle\partial\phi|\partial\phi\rangle
-\frac{1}{2} \sum_{\text{\scriptsize{points}}}e^{\langle\text{\scriptsize{points}}|\phi\rangle} (F_{\text{\scriptsize{points}}}^{(1)})^2
-\frac{1}{2} \sum_{\text{\scriptsize{quads}}}e^{\langle\text{\scriptsize{quads}}|\phi\rangle} ({ F}_{\text{\scriptsize{quads}}}^{(1)})^2
-\frac{1}{2} \sum_{\text{\scriptsize{lines}}} e^{\langle\text{\scriptsize{lines}}|\phi\rangle} ({ F}_{\text{\scriptsize{lines}}}^{(1)})^2\Bigg],
\ee
where in this case the sums run over the vectors listed in Table \ref{D=3FULL}.

\section{`Four Curious Supergravities' with $\mathcal{N}=1,2,4,8$ in $D=4$}\label{FOURCURIOUS}

In this final section we return to $D=4$ and demonstrate a natural series of Fano-plane-based truncations of $\N=8$ supergravity that yield the so-called `four curious supergravities' studied in \cite{PhysRevD.83.046007}. These theories can be summarised as
\begin{itemize}
\item $\N=8$ supergravity with coset $\frac{E_{7(7)}}{\text{SU}(8)}$,
\item $\N=4$ supergravity coupled to 6 vector multiplets with coset $\frac{\SL(2,\R)\times\SO(6,6)}{\SO(2)\times\SO(6)^2}$,
\item $\N=2$ supergravity coupled to 3 vector multiplets and 4 hyper multiplets with coset $\frac{\SL(2,\R)^3\times\SO(4,4)}{\SO(2)^3\times\SO(4)^2}$,
\item $\N=1$ supergravity coupled to 7 Wess-Zumino multiplets with coset $\frac{\SL(2,\R)^7}{\SO(2)^7}$.
\end{itemize}
Note that the truncations are rank-preserving; that is, each of the U-duality groups has rank 7. This means that in each case the root space is 7-dimensional, so we may continue to make use of imaginary octonions in our description. The truncations from $\N=8$ supergravity to the $\N=4,2,1$ theories can be carried out at the level of the root space using the Fano plane as a guide as follows:
\begin{itemize} 
\item For $\N=4$: choose one point of the Fano plane -- say $e_1$; take the unit Kirmse integers $\pm e_1$ to be the roots of $\SL(2,\R)$; the space of imaginary Kirmse integers orthogonal to $\pm e_1$ then corresponds to the root lattice of $\SO(6,6)$.
\item For $\N=2$: choose a line of the Fano plane -- say 124; take the unit Kirmse integers $\pm e_1$, $\pm e_2$ and $\pm e_4$ to be the roots of $\SL(2,\R)^3$; the space of imaginary Kirmse integers orthogonal to $\pm e_1$, $\pm e_2$ and $\pm e_4$  then corresponds to the root lattice of $\SO(4,4)$.
\item For $\N=1$: `choose' the whole Fano plane; take the unit Kirmse integers $\pm e_1,\pm e_2,\pm e_3,\pm e_4,\pm e_5\pm e_6,\pm e_7$ to be the roots of $\SL(2,\R)^7$.
\end{itemize}
In other words, the truncation series with $\N=8,4,2,1$ amounts to singling out 0, 1, 3 and 7 points of the Fano plane, respectively, retaining only the corresponding imaginary units and the space of roots orthogonal to these units. This is best illustrated by the example shown in Table \ref{CURIOUS}.

\begin{table}[ht]\vspace{0.5cm}
\begin{tabular}{c|c|c|c|c|c}

\hline\hline
& \multicolumn{2}{c|}{}& \multicolumn{2}{c|}{} &\\
$\N=8:~E_{7(7)}$
& \multicolumn{2}{c|}{$\N=4:~\SL(2,\R)\times\SO(6,6)$}& \multicolumn{2}{c|}{$\N=2:~\SL(2,\R)^3\times\SO(4,4)$} & $\N=1:~\SL(2,\R)^7$ \\  
& \multicolumn{2}{c|}{}& \multicolumn{2}{c|}{} & \\
\hline 
& $\SL(2,\R)$ & $\SO(6,6)$ &$\SL(2,\R)^3$& $\SO(4,4)$&\\ \cline{2-5}
$\pm e_1, \pm e_2, \pm e_3, \pm e_4,$ & $~~~~~~\pm e_1~~~~~~$ & $\pm e_2, \pm e_3, \pm e_4,$ &$~\pm e_1,\pm e_2,\pm e_4~$&$\pm e_3,$& $\pm e_1, \pm e_2, \pm e_3, \pm e_4,$\\ 
$\pm e_5,  \pm e_6, \pm e_7,$ &  & $\pm e_5,  \pm e_6, \pm e_7,$& &$\pm e_5,  \pm e_6, \pm e_7,$& $\pm e_5,  \pm e_6, \pm e_7$\\ 
$ \frac{1}{2}( \pm e_3 \pm  e_5 \pm e_6 \pm e_7),$ &  & $ \frac{1}{2}( \pm e_3 \pm  e_5 \pm e_6 \pm e_7),$ & & $\frac{1}{2}( \pm e_3 \pm  e_5 \pm e_6 \pm e_7)$& \\ 
$ \frac{1}{2}( \pm e_4 \pm  e_6 \pm e_7 \pm e_1),$ &  &&  &  & \\ 
$ \frac{1}{2}( \pm e_5 \pm  e_7 \pm e_1 \pm e_2),$ &  &&  &  & \\ 
$ \frac{1}{2}( \pm e_6 \pm  e_1 \pm e_2 \pm e_3),$ &  &&  &  & \\ 
$ \frac{1}{2}( \pm e_7 \pm  e_2 \pm e_3 \pm e_4),$ &  & $ \frac{1}{2}( \pm e_7 \pm  e_2 \pm e_3 \pm e_4),$&  &  & \\ 
$ \frac{1}{2}( \pm e_1 \pm  e_3 \pm e_4 \pm e_5),$ &  &&  &  & \\ 
$ \frac{1}{2}( \pm e_2 \pm  e_4 \pm e_5 \pm e_6)\,\,$ &  & $ \frac{1}{2}( \pm e_2 \pm  e_4 \pm e_5 \pm e_6)$&  &  & \\ 
&&&&\\
\hline\hline
\end{tabular}
\caption{$E_{7(7)}\supset \SL(2,\R)\times\SO(6,6)\supset\SL(2,\R)^3\times\SO(4,4)\supset\SL(2,\R)^7$ roots, corresponding to the four curious supergravities. The sets single out 0, 1, 3 and 7 points of the Fano plane, respectively, or equivalently 7, 3, 1 and 0 quadrangles.}\label{CURIOUS}
\end{table}

Counting the roots in each case (excluding $\SL(2,\R)$ factors) we have $(7\times2)+(7\times2^4)=126\rightarrow(6\times2)+(3\times2^4)=60\rightarrow(4\times2)+(1\times2^4)=24\rightarrow0$ for $E_{7(7)}\rightarrow{\SO}(6,6)\rightarrow{\SO}(4,4)$.

\section{Summary}

We have demonstrated how eleven-dimensional supergravity may be written over the octonions. The octonions are simply used in an alternative formulation of the usual Clifford algebra for the fermionic sector. However, the octonionic parameterisation leads to a new perspective in the bosonic sector upon dimensional reduction to the maximal supergravity theories in $D=4$ and $D=3$. 

In the $D=4$ case we write the seven coordinates of the internal dimensions as an imaginary octonion, leading us to interpret the seven dilatons as an imaginary octonion. The resulting parameterisation of the dilaton vectors (as imaginary Kirmse integers) in terms of the quadrangles of the Fano plane offers a simple way to interpret the sublattices corresponding to the four curious supergravities.

Upon reduction to $D=3$, the 240 dilaton vectors may be considered as the 240 unit octavian integers, and thus they form an algebra that is closed under multiplication. This is an interesting result in its own right, although what it can be used for -- or indeed whether it is useful at all -- is so far a mystery. We speculate that the algbera could have some utility in working with $D=3$ black hole solutions, in which dilaton vectors sometimes appear explicitly. However, for now we could call what we have found an \emph{answer without a question}.

\section{acknowledgements}

We would like to thank our supervisor Michael Duff for his guidance and for the initial inspiration for this paper. We thank Silvia Nagy for useful discussions and give a special thanks to Leron Borsten for his help with the dilaton vectors. The work of AA and MJH is supported by STFC studentships.

\begin{appendix}

\section{Complete List of Octonionic Dilaton Vectors}

In order to list the dilaton vectors in a concise form we introduce some additional terminology and notation. A \emph{flag} on the Fano plane $\F$ (or its dual $\widetilde{\F}$) is a pair $(ijk,i)$, consisting of an unoriented line $ijk$ and a point $i$ lying on that line. Since the line $ijk$ is unoriented, we write $\sigma(ijk)\in\bold{L}$ (or $\widetilde{\bold{L}}$), where $\sigma(ijk)$ is some permutation that puts $i,j$ and $k$ into the appropriate order.  There are $7\times3=21$ flags on the Fano plane (or its dual) and we denote the set of these as $\text{Fl}(\F)$ (or $\text{Fl}(\widetilde{\F})$). Note that any pair of distinct points $i,j$ on $\F$ (or $\widetilde{\F}$) uniquely defines a flag $(ijk,k)$, since choosing two points $i,j$ selects a unique line $ijk$, and giving preference to $i$ and $j$ over $k$ singles out $k$.

An \emph{anti-flag} is a pair $(ijk,l)$, consisting of an unoriented line $ijk$ and a point $l$ $\emph{not}$ lying on that line. There are $7\times4=28$ anti-flags and we denote the set of these as $\overline{\text{Fl}}(\F)$ (or $\overline{\text{Fl}}(\widetilde{\F})$ for its dual). Note that any triple of points $ijk$ that is \emph{not} a line on the Fano plane defines a unique anti-flag, since the compliment of that triple in the plane consists of four distinct points, three of which form a line.

Using \eqref{DVECTORS} one can compute the full set of dilaton vectors. Because of the parameterisations \eqref{D4F} and \eqref{D3F} the resulting vectors exhibit a correspondence with the Fano plane:
\begin{itemize}
\item $a_i$ carries a label $i$ corresponding to a point of $\widetilde{\F}$, which maps by duality to a line $ijk$ on $\F$; the complement of this line is a quadrangle $i'j'k'l'$ on $\F$; we find that $a_i=\tfrac{1}{2}(e_{i'}+e_{j'}+e_{k'}+e_{l'})$.
\item $a_{ij}$ ($i<j$) singles out a flag in $(ijk,k)\in{\text{Fl}}(\widetilde{\F})$ which maps to a flag $(lmn,l)\in\text{Fl}(\F)$; the resulting vector is $a_{ij}=\tfrac{1}{2}(e_l-e_m-e_n)$, with the different signs reflecting the flag $(lmn,l)$.
\item $a_{ijk},~ijk \in \widetilde{\bold{L}}$ clearly singles out the line $ijk$ on $\widetilde{\F}$, which maps to a point $i$ on ${\F}$; we find that $a_{ijk}=e_i$.
\item $a_{ijk},~ijk \notin \widetilde{\bold{L}}$ (with $~i<j<k$) corresponds to an anti-flag $A\in\overline{\text{Fl}}(\widetilde{\F})$, which maps to an anti-flag $(lmn,l')\in\overline{\text{Fl}}(\F)$; the complement of the unoriented line $lmn$ is an unoriented quadrangle $l'm'n'p'$ where the point $l'$ is distinguished by the flag $(lmn,l')$; the result is
$a_{ijk}=\tfrac{1}{2}(-e_{l'}+e_{m'}+e_{n'}+e_{p'})$.
\item $b_i$ corresponds to a point $i$ on $\Ftilde$, which gives a line $ijk$ on $\F$, giving $b_i=-\tfrac{1}{2}(e_i+e_j+e_k)$.
\item $b_{ij}$ ($i<j$) again selects a flag in $(ijk,k)\in{\text{Fl}}(\widetilde{\F})$ which maps to a flag $(lmn,l)\in\text{Fl}(\F)$; the complement of the unoriented line $lmn$ is an unoriented quadrangle $l'm'n'p'$, which is naturally split into two halves $l'm'$ and $n'p'$ by the flag $(lmn,l)$, since $l'$ and $m'$ lie on an unoriented line $ll'm'$ with $l$, while $n'$ and $p'$ lie on another line $ln'p'$ with $l$; the resulting vector is $b_{ij}=\tfrac{1}{2}(e_{l'}+e_{m'}-e_{n'}-e_{p'})$ with the overall sign dictated as follows: looking at the labels $i$ and $j$ of $b_{ij}$ we see that in general $i=n'$ and/or $j=l'$.
\end{itemize}

This is summarised in Table \ref{D=4FULL} and, since the $D=3$ case is very similar, we simply list its vectors in Table \ref{D=3FULL}.

\begin{table}[h!]
$\begin{array}{c|c|c|c}
\toprule
&&&\\
\text{Dilaton Vector}&\text{Fano Plane Interpretation}&\text{Octonionic Parameterisation}&\text{Number}\\ 
&&&\\
\hline
&&&\\
-a       & \text{Full Plane}&\tfrac{1}{2}(e_1+e_2+e_3+e_4+e_5+e_6+e_7)  &1\\
&&&\\
 -a_{i}&\text{Point $i$ on $\widetilde{\F}$}~\leftrightarrow~\text{Line $ijk$ on ${\F}$}&\tfrac{1}{2}(e_{i'}+e_{j'}+e_{k'}+e_{l'}),~ i'j'k'l' \in \bold{Q}&7\\
&&\\
a_{ij},~i<j&(ijk,k)\in{\text{Fl}}(\widetilde{\F})~\leftrightarrow~(lmn,l)\in\text{Fl}(\F)&\tfrac{1}{2}(e_l-e_m-e_n),~\sigma(lmn)\in {\bold{L}} &21\\
&&\\
a_{ijk},~ijk \in \widetilde{\bold{L}}&\text{Line $ijk$ on $\widetilde{\F}$}~\leftrightarrow~\text{Point $i$ on ${\F}$}& e_i&7\\
&&\\
a_{ijk},~i<j<k,~ijk \notin \widetilde{\bold{L}}&A\in\overline{\text{Fl}}(\widetilde{\F})~\leftrightarrow~(lmn,l')\in\overline{\text{Fl}}(\F)& 
\tfrac{1}{2}(-e_{l'}+e_{m'}+e_{n'}+e_{p'}),~ \sigma(l'm'n'p') \in \bold{Q} &28\\
&&&\\
b_{i}&\text{Point $i$ on $\widetilde{\F}$}~\leftrightarrow~\text{Line $ijk$ on ${\F}$}&-\tfrac{1}{2}(e_i+e_j+e_k),~ijk\in\bold{L}&7\\
&&\\
b_{ij},~i<j &(ijk,k)\in{\text{Fl}}(\widetilde{\F})~\leftrightarrow~(lmn,l)\in\text{Fl}(\F)&\tfrac{1}{2}(e_{l'}+e_{m'}-e_{n'}-e_{p'}),~ \sigma(l'm'n'p') \in \bold{Q}, &21\\
&&\sigma(ll'm'),\sigma(ln'p')\in\bold{L},~ \text{$i=n'$ and/or $j=l'$}&\\

&&&\\
 \hline
\hline
\end{array}$
\caption{Complete list of the octonionic $D=4$ dilaton vectors. The vectors (or Kirmse integers) $ {a}_{ijk}$, $ {b}_{ij}$ and $- {a}_i$ are the positive roots of $E_{7(7)}$, while $ {a}_{ij}$ and $ {b}_i$ make up the positive weights of the $\bold{56}$ representation. We use the notation $\sigma(lmn)\in {\bold{L}}$ to mean that we can find some permutation of $lmn$ that gives a line in $\bold{L}$ (we write this because, strictly speaking, the lines in $\bold{L}$ consist of ordered triples of points).}\label{D=4FULL}
\end{table}

\begin{table}[h!]
$\begin{array}{c|c|c|c}
\toprule
&&&\\
\text{Dilaton Vector}&\text{Fano Plane Interpretation}&\text{Octonionic Parameterisation}&\text{Number}\\ 
&&&\\
\hline
&&&\\
-b_0       & \text{--}&1  &1\\
&&&\\
 -b_{i}&\text{Point $i$ on $\widetilde{\F}$}~\leftrightarrow~\text{Line $ijk$ on ${\F}$}&\tfrac{1}{2}(1+e_{i}+e_{j}+e_{k}),~ ijk \in \bold{L}&7\\
&&\\
b_{0i}&\text{Point $i$ on $\widetilde{\F}$}~\leftrightarrow~\text{Line $ijk$ on ${\F}$}&\tfrac{1}{2}(-1+e_{i}+e_{j}+e_{k}),~ ijk \in \bold{L}&7\\
&&\\
b_{ij},~i<j &(ijk,k)\in{\text{Fl}}(\widetilde{\F})~\leftrightarrow~(lmn,l)\in\text{Fl}(\F)&\tfrac{1}{2}(e_{l'}+e_{m'}-e_{n'}-e_{p'}),~ \sigma(l'm'n'p') \in \bold{Q}, &21\\
&&\sigma(ll'm'),\sigma(ln'p')\in\bold{L},~ \text{$i=n'$ and/or $j=l'$}&\\
&&&\\
a_{0i}&\text{Point $i$ on $\widetilde{\F}$}~\leftrightarrow~\text{Line $ijk$ on ${\F}$}&\tfrac{1}{2}(e_{i'}+e_{j'}+e_{k'}+e_{l'}),~ i'j'k'l' \in \bold{Q}&7\\
&&\\
a_{ij},~i<j&(ijk,k)\in{\text{Fl}}(\widetilde{\F})~\leftrightarrow~(lmn,l)\in\text{Fl}(\F)&\tfrac{1}{2}(-1+e_l-e_m-e_n),~\sigma(lmn)\in {\bold{L}} &21\\
&&\\
a_{ij0},~i<j&(ijk,k)\in{\text{Fl}}(\widetilde{\F})~\leftrightarrow~(lmn,l)\in\text{Fl}(\F)&\tfrac{1}{2}(+1+e_l-e_m-e_n),~\sigma(lmn)\in {\bold{L}} &21\\
&&\\
a_{ijk},~ijk \in \widetilde{\bold{L}}&\text{Line $ijk$ on $\widetilde{\F}$}~\leftrightarrow~\text{Point $i$ on ${\F}$}& e_i&7\\
&&\\
a_{ijk},~i<j<k,~ijk \notin \widetilde{\bold{L}}&A\in\overline{\text{Fl}}(\widetilde{\F})~\leftrightarrow~(lmn,l')\in\overline{\text{Fl}}(\F)& 
\tfrac{1}{2}(-e_{l'}+e_{m'}+e_{n'}+e_{p'}),~ \sigma(l'm'n'p') \in \bold{Q} &28\\
&&&\\
 \hline
\hline
\end{array}$
\caption{Complete list of the $D=3$ dilaton vectors ($- {a}_{ab}$, $ {a}_{abc}$, $- {b}_{a}$ and $ {b}_{ab}$) written as Kirmse integers. Together all the dilaton vectors make up the positive roots of $E_{8(8)}$. }\label{D=3FULL}
\end{table}

\end{appendix}

\bibliography{Octonions}

\end{document}